\documentstyle[preprint,tighten,aps,floats,epsfig]{revtex}
\def\slashed{{/}\mskip-10.0mu}
\def\pcircslash{\slashed {p\mskip -5mu ^{^\circ}}}
\begin{document}
\draft
\vskip 2cm

\title{The Critical Hopping Parameter in ${\cal O}(a)$ improved Lattice QCD}

\author{H. Panagopoulos and Y. Proestos\footnote{Present address:
Department of Physics, Ohio State University, Columbus, OH 43210, USA}}
\address{Department of Physics, University of Cyprus, P.O. Box 20537,
Nicosia CY-1678, Cyprus \\
{\it email: }{\tt haris@ucy.ac.cy, yiannis@mps.ohio-state.edu}}
\vskip 3mm

\date{\today}

\maketitle

\begin{abstract}

We calculate the critical value of the hopping
parameter, $\kappa_c$, in ${\cal O}(a)$ improved Lattice QCD, to two
loops in perturbation theory. We employ the Sheikholeslami-Wohlert
(clover) improved action for Wilson fermions.

The quantity which we study is a typical case of a vacuum expectation
value resulting in an additive renormalization; as such, it is
characterized by a power (linear) divergence in the lattice spacing, and
its calculation lies at the limits of applicability of perturbation theory. 

The dependence of our results  on the number of colors $N$, the number of
fermionic flavors $N_f$, and the clover parameter $c_{\rm SW}$, is
shown explicitly. 
We compare our results to non perturbative evaluations of $\kappa_c$
coming from Monte Carlo simulations.


\end{abstract}

\newpage


\section{Introduction}
\label{introduction}

In this paper we calculate the critical value of the hopping
parameter $\kappa_c$ in Lattice QCD, to two
loops in perturbation theory. We employ the ${\cal O}(a)$ improved
Sheikholeslami-Wohlert~\cite{SW} (clover) action for Wilson fermions;
this action is widely used nowadays in Monte Carlo simulations, as a means
of reducing finite lattice spacing effects, leading to a faster
approach to the continuum.

The Wilson fermionic action is a standard implementation of fermions
on the lattice. It circumvents the notorious doubling problem by means of a
higher derivative term, 
which removes unphysical propagator poles and has a vanishing
classical continuum limit; at the same time, the action is 
strictly local, which is very advantageous for numerical simulation.
The price one pays for strict locality and absence of doublers is, of
course, well known: The higher derivative term breaks chiral
invariance explicitly. Thus, merely setting the bare fermionic mass to
zero is not sufficient to ensure chiral symmetry in the quantum
continuum limit; quantum corrections introduce an additive
renormalization to the fermionic mass, which must then be fine tuned
to have a vanishing renormalized value. Consequently, the
hopping parameter $\kappa$, which is very simply related to the fermion mass,
must be appropriately shifted from its naive value, in order to recover chiral
invariance.

By dimensional power counting, the additive mass renormalization is
seen to be linearly divergent with the lattice spacing. This adverse 
feature of Wilson fermions, typical of vacuum expectation values of
local objects,
poses an additional problem to a 
perturbative treatment, aside from the usual issues related to lack
of Borel summability. Indeed, our calculation serves as a
check on the limits of applicability of perturbation theory, by
comparison with non perturbative results coming from Monte Carlo simulations.

In the present work we will follow the procedure and notation of
Ref.~\cite{FP}, in which $\kappa_c$ 
was computed using the Wilson fermionic action without
${\cal O}(a)$ improvement. The results of Ref.~\cite{FP} were
recently confirmed in Ref.~\cite{CPR}, in which a coordinate space
method was used to achieve even greater accuracy.

The critical fermionic mass and hopping
parameter will now depend not only on the number of colors $N$
and of fermionic flavours $N_f$, but also on the
free parameter $c_{\rm SW}$ which appears in the clover action (see
next Section); we will keep this dependence explicit in our results.

In Sec.~\ref{sec2} we define the quantities which we set out to
compute, and describe our calculation.
In Sec.~\ref{sec3} we present our results and compare with Monte Carlo
evaluations. Finally, in Sec.~\ref{sec4} we obtain improved
estimates coming from a tadpole resummation. 

\section{Formulation of the problem}
\label{sec2}

Our starting point is the Wilson formulation of the QCD action on the
lattice, with the addition of the clover (SW)~\cite{SW} fermion
term. Its action reads, in standard notation:

\begin{eqnarray}
S_L &=&  {1\over g_0^2} \sum_{x,\,\mu,\,\nu}
{\rm Tr}\left[ 1 - U_{\mu,\,\nu}(x) \right]  +
\sum_{f}\sum_{x} (4r+m_B)\bar{\psi}_{f}(x)\psi_f(x) 
\nonumber \\
&-&{1\over 2}\sum_{f}\sum_{x,\,\mu}
\left[ 
\bar{\psi}_{f}(x)\left( r - \gamma_\mu\right)
U_{\mu}(x)\psi_f(x+\hat{\mu})+
\bar{\psi}_f(x+\hat{\mu})\left( r + \gamma_\mu\right)
U_{\mu}(x)^\dagger
\psi_{f}(x)\right]\nonumber \\
&+& {i\over 4}\,c_{\rm SW}\,\sum_{f}\sum_{x,\,\mu,\,\nu} \bar{\psi}_{f}(x)
\sigma_{\mu\nu} {\hat F}_{\mu\nu}(x) \psi_f(x),
\label{latact}
\end{eqnarray}
\begin{equation}
{\rm where:}\qquad {\hat F}_{\mu\nu} \equiv {1\over8}\,
(Q_{\mu\nu} - Q_{\nu\mu}), \qquad 
Q_{\mu\nu} = U_{\mu,\,\nu} + U_{\nu,\,{-}\mu} + U_{{-}\mu,\,{-}\nu} + U_{{-}\nu,\,\mu}
\end{equation}
Here $U_{\mu,\,\nu}(x)$ is the usual product of link variables
$U_{\mu}(x)$ along the perimeter of a plaquette in the $\mu$-$\nu$
directions, originating at $x$;
$g_0$ denotes the bare coupling constant; $r$ is the Wilson parameter;
$f$ is a flavor index; $\sigma_{\mu\nu} =
(i/2) [\gamma_\mu,\,\gamma_\nu]$. 
Powers of the lattice spacing $a$ have been omitted and 
may be directly reinserted by dimensional counting.

We use the standard covariant gauge-fixing term; in terms of
the vector field $Q_\mu(x)$ $\left[U_{\mu}(x)= \exp(i\,g_0\,Q_\mu(x))\right]$, it
reads:
\begin{equation}
S_{gf} = \lambda_0 \sum_{\mu , \nu} \sum_{x}
\hbox{Tr} \, \Delta^-_{\mu} Q_{\mu}(x) \Delta^-_{\nu} Q_{\nu}(x), \qquad
\Delta^-_{\mu} Q_{\nu}(x) \equiv Q_{\nu}(x - {\hat \mu}) - Q_{\nu}(x).
\end{equation}
Having to compute a gauge invariant quantity, we chose to work
in the Feynman gauge, $\lambda_0 = 1$. 
Covariant gauge fixing produces the following
action for the ghost fields $\omega$ and $\overline\omega$
\begin{eqnarray}
S_{gh} &=& 2 \sum_{x} \sum_{\mu} \hbox{Tr} \,
(\Delta^+_{\mu}\omega(x))^{\dagger} \Bigl( \Delta^+_{\mu}\omega(x) +
i g_0 \left[Q_{\mu}(x),
\omega(x)\right] + \case{1}{2}
i g_0 \left[Q_{\mu}(x), \Delta^+_{\mu}\omega(x) \right] \nonumber\\
 & & \quad - \case{1}{12}
g_0^2 \left[Q_{\mu}(x), \left[ Q_{\mu}(x),
\Delta^+_{\mu}\omega(x)\right]\right] + \cdots \Bigr), \qquad
\Delta^+_{\mu}\omega(x) \equiv \omega(x + {\hat \mu}) - \omega(x).
\end{eqnarray}
Finally the change of integration variables from links to vector
fields yields a jacobian that can be rewritten as 
the usual measure term $S_m$ in the action:
\begin{equation}
S_{m} = \frac{1}{12} N g_0^2 \sum_{x} \sum_{\mu} \hbox{Tr} \,
Q_{\mu}(x) Q_{\mu}(x) + \cdots
\end{equation}
In $S_{gh}$ and $S_m$ we have written out only
terms relevant to our computation.
The full action is: $S = S_L + S_{gf} + S_{gh} + S_m.$

The bare fermionic mass $m_B$ must be set to zero for
chiral invariance in the classical continuum limit. The value of the
parameter $c_{\rm SW}$ can be chosen arbitrarily; it is normally 
tuned in a way as to minimize ${\cal O}(a)$ effects.
Terms proportional to $r$ in the action, as well as the clover terms,
break chiral invariance. They vanish in the classical continuum
limit; at the quantum level, they induce nonvanishing,
flavor-independent corrections to the fermion masses.
Numerical simulation algorithms usually employ the hopping parameter, 
\begin{equation}
\kappa\equiv{1\over 2\,m_B\,a + 8\,r}
\end{equation}
as an adjustable quantity. Its critical value, at which chiral symmetry
is restored, is thus $1/8r$ classically, but gets shifted by quantum
effects.

The renormalized mass can be calculated in textbook fashion from the
fermion self--energy. Denoting by $\Sigma^L(p,m_B,g_0)$ the truncated,
one particle irreducible fermionic two-point function, we have for the
fermionic propagator:
\begin{eqnarray}
S(p)&=& \left[ i \,\pcircslash + m(p)- \Sigma^L(p,m_B,g_0)\right]^{-1}\\
{\rm where:}\qquad \pcircslash &=& \sum_\mu\gamma_\mu {1\over a} \sin(ap^\mu), \quad m(p) =
m_B + {2r\over a} \sum_\mu \sin^2(ap^\mu/2).\nonumber
\end{eqnarray}

To restore the explicit breaking of chiral invariance, we require that the renormalized mass vanish:
\begin{equation}
S^{-1}(0) = 0 \qquad\Longrightarrow\qquad m_B = \Sigma^L(0,m_B,g_0)
\end{equation}
The above is a recursive equation for $m_B$, which can be solved order
by order in perturbation theory.

\bigskip
We write the loop expansion of $\Sigma^L$ as:

\begin{equation}
\Sigma^L(0,m_B,g_0) = g_0^2 \, \Sigma^{(1)} + g_0^4 \, \Sigma^{(2)} + \cdots
\end{equation}

Two diagrams contribute to $\Sigma^{(1)}$, shown in Fig. I. In these
diagrams, the fermion mass 
must be set to its tree level value, $m_B\to 0$. 

\bigskip
\hskip4.0cm\psfig{figure=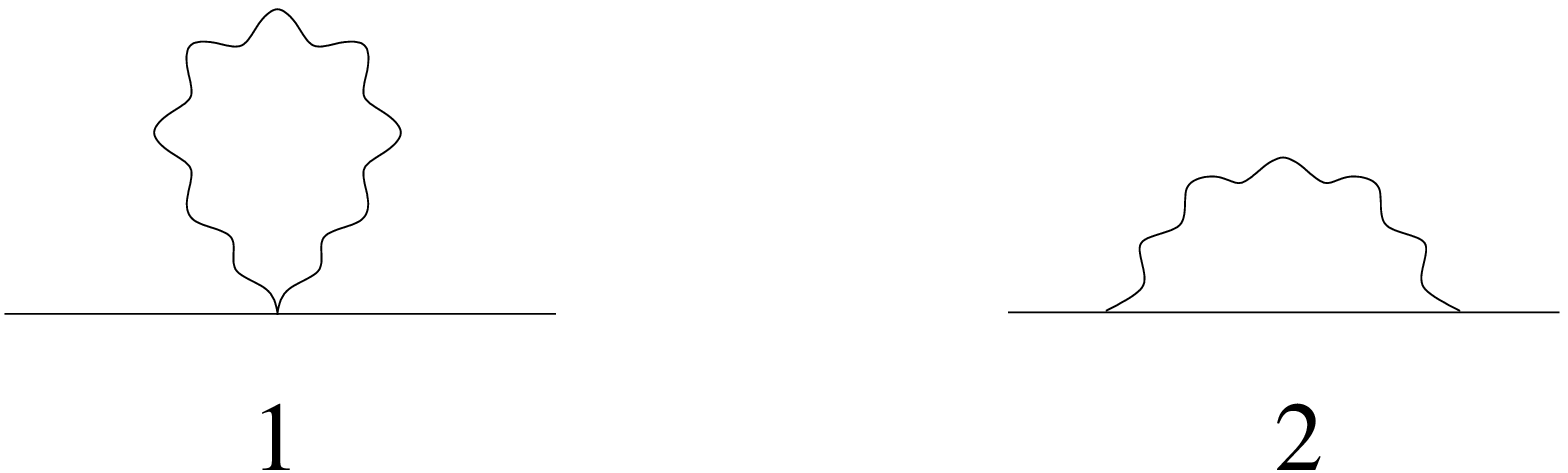,height=2truecm}\hskip1.0cm
\nopagebreak\medskip

\noindent
{\small FIGURE I.\ \ One-loop diagrams contributing to $\Sigma^L$.
Wavy (solid) lines represent gluons (fermions).}

\bigskip
The quantity $\Sigma^{(2)}$ receives contributions from 
a total of 26 diagrams, shown in 
Fig. II. Genuine 2-loop diagrams must again be evaluated at
$m_B\to 0$; in addition, one must include to this order the 1-loop diagram
containing an ${\cal O}(g_0^2)$ mass counterterm (diagram 23).

The contribution of the $i^{\rm th}$ diagram 
can be written in the form

\begin{equation}
(N^2 - 1) \cdot \sum_{j=0}^4 \left(c^{(j)}_{1,i} + \frac{c^{(j)}_{2,i}}{N^2} + 
\frac{N_f}{N} \, c^{(j)}_{3,i}\right) \, {c_{\rm SW}}^j
\label{c4}
\end{equation}
\noindent
where $c^{(j)}_{1,i},c^{(j)}_{2,i},c^{(j)}_{3,i}$ are numerical
constants. The dependence on $c_{\rm SW}$ is seen to be polynomial of
degree 4, as can be verified by inspection of Fig. II.

 Certain sets of diagrams, corresponding to renormalization of loop
 propagators, must be evaluated together in
order to obtain an infrared-convergent result: These are diagrams 7+8+9+10+11,
12+13, 14+15+16+17+18, 19+20, 21+22+23.

\medskip
\hskip2.0cm\psfig{figure=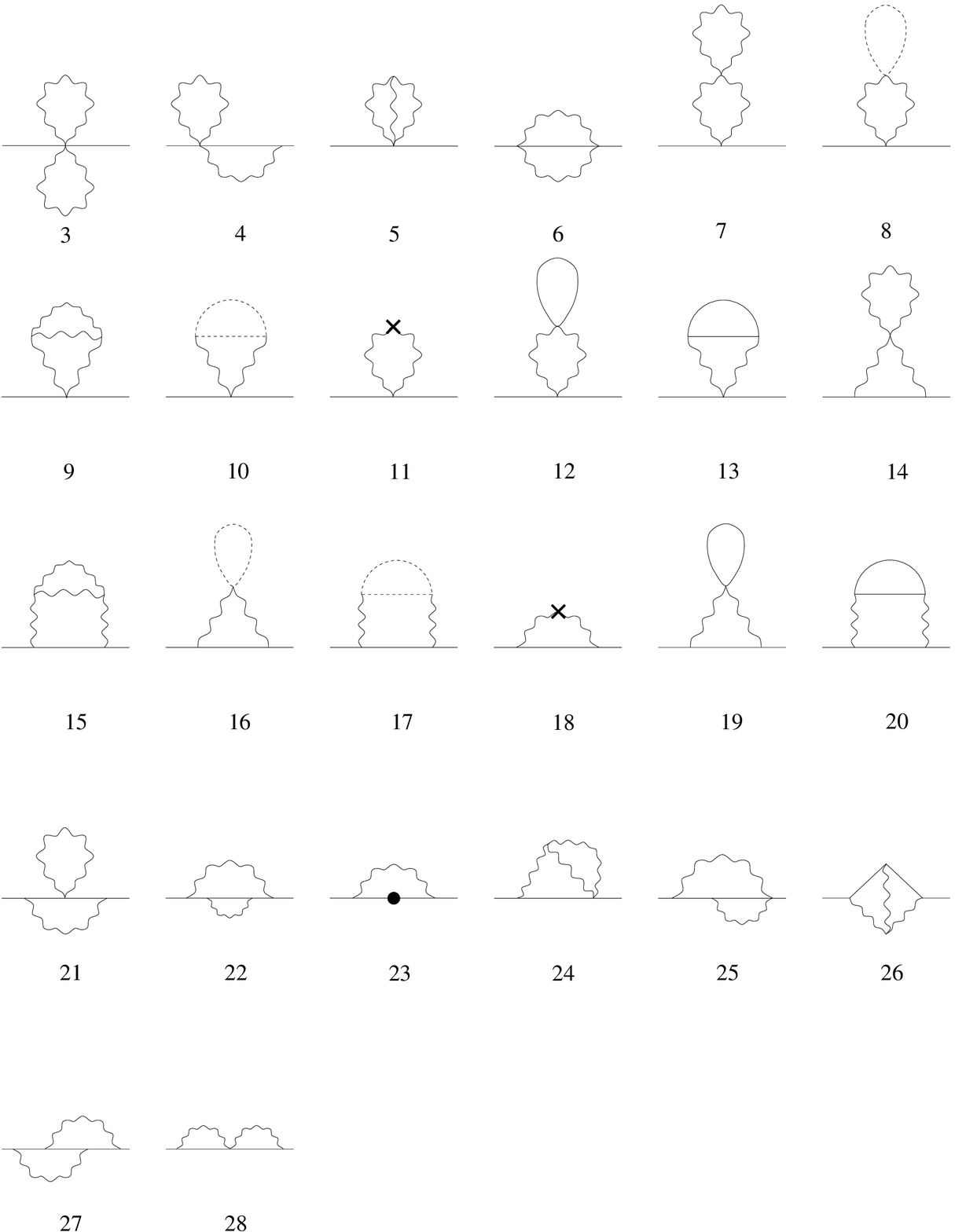,height =14truecm}\hskip1.0cm
\bigskip
\nobreak

\noindent
{\small FIGURE II.\ \ Two-loop diagrams contributing to $\Sigma^L$.
Wavy (solid, dotted) lines represent gluons (fermions,
ghosts). Crosses denote vertices stemming from the measure part of the
action; a solid circle is a fermion mass counterterm.}

\section{Numerical Results}
\label{sec3}

Evaluating the two diagrams of Figure I, we find for $\Sigma^{(1)}$ : 
\begin{equation}
\begin{array}{rllll}
\Sigma^{(1)}= {\displaystyle N^2-1\over \displaystyle N} \,
\bigl(&-0.15493339023106021 &&&{\rm (diagram\ 1)}\\
      &-0.00792366847979(1)&+c_{\rm SW}    &0.04348303388205(10)&\\
      &                    &+{c_{\rm SW}}^2&0.01809576878142(1)\ \ \bigr)\quad
                         &{\rm (diagram\ 2)}.
\end{array}
\label{Sigma1}
\end{equation}
Here and below we set $r$ to its usual value, $r=1\,$. One- and
two-loop results pertaining to $c_{\rm SW} = 0$ are as in
Ref.~\cite{FP}, and can be found with greater accuracy in
Ref.~\cite{CPR}.

For $c_{\rm SW}\ne 0$, only one-loop results exist so far in the
literature; a recent presentation for the case $c_{\rm SW}{=}1$ (see
Ref.~\cite{BWW} and earlier references therein) is in perfect
agreement with our Eq.~(\ref{Sigma1}):
\begin{equation}
\begin{array}{rll}
\Sigma^{(1)}(N{=}3, c_{\rm SW}{=}1) = &-0.2700753495(2) \qquad\qquad&
{\rm [Ref.~\cite{BWW}]}\\
                                      &-0.2700753494597(5) &
{\rm [The\ present\ work,\ Eq.~(\ref{Sigma1})]}.
\end{array}
\end{equation}

It should be clear to the reader that such a high level of precision
is hardly relevant {\it per se}, especially given the expected
deviation from nonperturbative results; nevertheless, it serves as a
testing ground for both accuracy and efficiency of our methods, in
view of the more demanding higher loop 
calculations. Further, high precision is called for in the context of
the Schr\"odinger functional computation, to permit a stable
extrapolation of various 
parameters to infinite lattice (see Ref.~\cite{BWW}). Regarding
efficiency, let us note that the numerical integrations leading to
Eq.~(\ref{Sigma1}) require a mere $\sim 10$ {\tt min} of CPU time on a
typical 1GHz Pentium III processor.

\medskip
We now turn to the much more cumbersome evaluation of the two-loop
diagrams, which is the crux of our present computation. As in
Ref.~\cite{FP}, we use a Mathematica package which we
have developed for symbolic manipulations in lattice perturbation
theory (see, e.g., Ref. \cite{CFPV}). For the purposes of the
present work (and a related work on the $\beta$-function~\cite{BP}),
we have augmented the package to include the vertices of the clover action.

In Tables I, II, III and IV we present the values of the coefficients
$c^{(1)}_{k,i},\ c^{(2)}_{k,i},\ c^{(3)}_{k,i},\ c^{(4)}_{k,i}$,
respectively. The ${\cal O}({c_{\rm SW}}^0)$ coefficients $c^{(0)}_{k,i}$
are as in Ref.~\cite{FP}, and have been listed in Table V for
completeness. Diagrams giving vanishing contributions to a given power
of $c_{\rm SW}$ have been omitted from the corresponding table.

The momentum integrations leading to the
values of each coefficient are performed
numerically on lattices of varying size $L\le 32$, and then
extrapolated to infinite lattice size using a broad spectrum
of functional forms of the type: $\sum_{i,j} e_{ij} (\ln L)^j/L^i$.
The systematic error resulting from the extrapolations has been estimated
rather conservatively using the procedure of Ref.~\cite{CFPV}, and has
been included in the tables. 

One important consistency check can be performed on those diagrams
which are separately IR divergent; taken together in groups, as listed
at the end of Section II, they give finite and very stable
extrapolations for the coefficients of each power of $c_{\rm SW}$.
Several other consistency checks stem from exact relations among
various coefficients; to name a few: 
\begin{equation}
\begin{array}{lll}
  c_{2,4}^{(1)}  &=& \phantom{-}b_2^{(1)}\, (1/2 - b_1^{(0)})\,/\,4 \\
  c_{2,4}^{(2)}  &=& \phantom{-}b_2^{(2)} /\,4 \\
  c_{2,7}^{(0)}  &=& -b_1^{(0)}\, 3\,/\,32 \\
  c_{2,14}^{(1)} &=& - b_2^{(1)} /\,8 \\
  c_{2,14}^{(2)} &=& - b_2^{(2)} /\,8   
\end{array}
\label{consistency}
\end{equation}                           
Here, $b_i^{(j)}$ are the coefficients of the $i$th 1-loop diagram,
multiplying ${c_{\rm SW}}^j$, as displayed in Eq.~(\ref{Sigma1}).
Comparing with our numerical values of Tables I-IV, we find agreement
well within the error bars.

Leaving the choice of values for $N$, $N_f$ and $c_{\rm SW}$
unspecified, our result takes the form:
\begin{equation}
\begin{array}{rllll}
\Sigma^{(2)} = (N^2 {-} 1)\bigl[
 {\phantom{+}}&
   (-0.017537(3)   &+1/N^2\ 0.016567(2)   &+N_f/N\ 0.00118618(8) ) &   \\
 {+}&(\phantom{+}
     0.002601(2)   &-1/N^2\ 0.0005597(7)  &-N_f/N\ 0.0005459(2) )  &c_{\rm SW} \\
 {+}&(-0.0001556(3)  &+1/N^2\ 0.0026226(2)  &+N_f/N\ 0.0013652(1) )  &{c_{\rm SW}}^2\\
 {+}&(-0.00016315(6) &+1/N^2\ 0.00015803(6) &-N_f/N\ 0.00069225(3) ) &{c_{\rm SW}}^3\\
 {+}&(-0.000017219(2)&+1/N^2\ 0.000042829(3)&-N_f/N\ 0.000198100(7) )&{c_{\rm SW}}^4 \bigr]
\label{Sigma2}
\end{array}
\end{equation}

To make more direct contact with non-perturbative results, we evaluate
$\Sigma^{(2)}$ at $N=3$ and $N_f= 0,2$, obtaining:
\begin{equation}
\begin{array}{rlllll}
\Sigma^{(2)}(N=3, N_f=0) = 
 &-0.12557(3) &+0.02031(2) &c_{\rm SW}    &+0.001087(3)  &{c_{\rm SW}}^2\\
 &            &-0.001165(1)&{c_{\rm SW}}^3&-0.00009968(2)&{c_{\rm SW}}^4\\[0.5ex]
\Sigma^{(2)}(N=3, N_f=2) = 
 &-0.11924(3) &+0.01740(2) &c_{\rm SW}    &+0.008368(3)  &{c_{\rm SW}}^2\\
 &            &-0.004857(1)&{c_{\rm SW}}^3&-0.0011562(1) &{c_{\rm SW}}^4
\label{Sigma2n3}
\end{array}
\end{equation}
Eqs.~(\ref{Sigma1}, \ref{Sigma2n3}) lead immediately to the 1- and
2-loop results for the critical mass: $m_c^{(1)} =
g_0^2 \Sigma^{(1)}$, 
$m_c^{(2)} = g_0^2 \Sigma^{(1)} + g_0^4
\Sigma^{(2)}$, and the corresponding hopping parameter
$\kappa_c = 1/(2\, m_c \, a + 8\, r)$.

A number of non-perturbative determinations of $\kappa_c$ 
exist in the literature for particular values of 
$\beta = 2N/g_0^2$ and $c_{\rm SW}\ne 0$, see
e.g. Refs.~\cite{LSSWW,Bowler} (quenched case) and
Refs.~\cite{JS,UKQCD} (unquenched, $N_f=2$). We present these in Table
VI, together with the 1- and 2-loop results ($\kappa_c^{(1)}$,
$\kappa_c^{(2)}$).
Also included in the Table are the
improved results obtained with the method described in the following
Section.

\section{Improved Perturbation Theory}
\label{sec4}

In order to obtain improved estimates from lattice perturbation
theory, one may perform a resummation to all orders of the so-called
``cactus'' diagrams~\cite{cactus1,cactus2,cactus3}. 
Briefly stated, these
are gauge--invariant tadpole diagrams which become disconnected if any one of their
vertices is removed. The original motivation of this procedure is the well known
observation of ``tadpole dominance'' in lattice perturbation theory.
In the following we adapt the calculation of Ref.~\cite{FP} to the
clover action. We refer to Ref.~\cite{cactus1} for definitions and
analytical results. 

Since the contribution of standard tadpole diagrams is not gauge invariant,
the class of gauge invariant diagrams we are considering needs further specification.
By the Baker-Campbell-Hausdorff (BCH) formula, the product of link variables
along the perimeter of a plaquette can be written as
\begin{eqnarray}
U_{x,\mu\nu}&& = 
e^{i g_0 A_{x,\mu}} e^{i g_0 A_{x+\mu,\nu}} e^{-i
g_0 A_{x+\nu,\mu}} e^{-i g_0 A_{x,\nu}} \nonumber \\
&&=\exp\left\{i g_0 (A_{x,\mu} + A_{x+\mu,\nu} -
A_{x+\nu,\mu} - A_{x,\nu}) + {\cal O}(g_0^2) \right\} \nonumber \\
&&=\exp\left\{ i g_0 F_{x,\mu\nu}^{(1)} +  i g_0^2
F_{x,\mu\nu}^{(2)} + {\cal O}(g_0^4)
\right\}
\end{eqnarray}
The diagrams that we propose to resum to all orders are the cactus
diagrams made of vertices containing $F_{x,\mu\nu}^{(1)}\,$.
Terms of this type come from the pure gluon  part of
the lattice action.  These diagrams dress the transverse gluon
propagator $P_A$ leading to an improved propagator $P_A^{(I)}$,
which is a multiple of the bare transverse one:
\begin{equation}
P_A^{(I)} = {P_A\over 1-w(g_0)},
\label{propdr}
\end{equation}
where the factor $w(g_0)$ will depend on $g_0$ and
$N$, but not on the momentum.
The function $w(g_0)$ can be extracted by an
appropriate algebraic equation that
 has been derived in Ref.~\cite{cactus1} and that can be easily
solved numerically; for $SU(3)$, $w(g_0)$ satisfies:
\begin{equation}
u \, e^{-u/3} \, \left[u^2 /3 - 4u +8\right]  = 2 g_0^2, \qquad 
u(g_0) \equiv {g_0^2 \over 4 (1-w(g_0))}.
\label{w}
\end{equation}
The vertices coming from the gluon part of the action, Eq.~(\ref{latact}),
get also dressed using a procedure similar to the one leading to 
Eq.~(\ref{propdr}) \cite{cactus1}.
Vertices coming from the Wilson part of the fermionic action stay
unchanged, since their  
definition contains no plaquettes on which to apply the linear BCH
formula; the 3- and 4-point vertices of the clover action, on the
other hand, acquire
simply a factor of $(1-w(g_0))$ \cite{cactus2}.

One can apply  the resummation of cactus diagrams to the calculation of
additive and multiplicative renormalizations of lattice operators.
Applied to a number of cases of interest~\cite{cactus1,cactus2}, this procedure yields 
remarkable improvements when compared with the available nonperturbative estimates.
As regards numerical comparison with other improvement schemes
(tadpole improvement, boosted perturbation theory,
etc.)~\cite{Parisi-81,L-M-93}, cactus  
resummation fares equally well on all the cases
studied~\cite{cactus3}. 

One advantageous feature of cactus resummation, in comparison to other
schemes of improved perturbation theory, is the possibility of
systematically incorporating higher loop diagrams. The present
calculation exemplifies this feature, as we will now show.

Dressing the 1-loop results is quite straightforward: the fermionic
propagator stays unchanged, the gluon propagator
gets multiplied by $1/(1-w(g_0)$ and the dressing of the fermionic
vertices amounts to a rescaling: $c_{\rm SW} \rightarrow c_{\rm SW}\,
(1-w(g_0))$. The resulting values, 
$\kappa^{(1)}_{c,\,\rm dressed}$, are
shown in Table \ref{tab6}. It is worth
noting that these values already fare better than the much more
laborious undressed 2-loop results.

We now turn to dressing the 2-loop results. Here, one must take care
to avoid double counting: A part of diagrams 4, 7 and 14 has already been
included in dressing the 1-loop result, and must be explicitly
subtracted from $\Sigma^{(2)}$ before dressing. Fortunately, this part
(we shall denote it by $\Sigma^{(2)}_{\rm sub}$)
is easy to identify, as it necessarily includes all of the $1/N^2$ part
in diagrams 7, 14, and the $1/N^2$ part of diagram 4 involving a clover
5-point vertex. A simple exercise in contraction of $SU(N)$
generators shows that $\Sigma^{(2)}_{\rm sub}$ is proportional to
$(2N^2-3)(N^2-1)/(3N^2)$. There follows without difficulty that:
\begin{eqnarray}
\Sigma^{(2)}_{\rm sub} &=& -(2N^2-3) (N^2-1)/(3N^2) \cdot\nonumber\\
      &&\qquad [b_2^{(1)} c_{\rm SW}/8 + c_{2,4}^{(2)}\, {c_{\rm SW}}^2 +
       c_{2,7}^{(0)} + 
       c_{2,14}^{(0)} + c_{2,14}^{(1)}\, c_{\rm SW} + c_{2,14}^{(2)}\, {c_{\rm SW}}^2]
\end{eqnarray}
(cf. Eq. (\ref{consistency})).

A potential complication is presented by gluon vertices. While the
3-gluon vertex dresses by a mere factor of $(1-w(g_0))$, the dressed
4-gluon vertex contains a term which is not simply a multiple of its
bare counterpart (see Appendix C of Ref.~\cite{cactus1}). 
It is easy to check, however, that this term
must simply be dropped, being precisely
the one which has already been taken into account in dressing the
1-loop result; the 
remainder dresses in the same way as the 3-gluon vertex. 
The very same situation prevails with the dressed 5-point vertex of
the clover action, as in diagram 4.

In conclusion, cactus resummation applied to the 2-loop quantity
$\Sigma^{(2)}$ leads to the following rather simple recipe:
\begin{equation}
m^{(2)}_{c,\,\rm dressed} = \Sigma^{(1)}\, {g_0^2\over 1-w(g_0)} + 
(\Sigma^{(2)}- \Sigma^{(2)}_{\rm sub})\, {g_0^4\over [1-w(g_0)]^2}\ 
\Biggr|_{\ c_{\rm SW}\,\rightarrow \,c_{\rm SW} (1-w(g_0))}
\label{m2dressed}
\end{equation}
(For the values of $\beta$ in Table \ref{tab6},
$\beta =
5.20, 5.26, 5.29, 5.7, 6.0, 6.2, 12.0, 24.0$,
we obtain from
Eq.~(\ref{w}): $1-w(g_0) 
= 0.697146$, 0.701957, 0.704298, 0.732579, 0.749775, 0.759969,
0.887765, 0.946087, respectively.) 

\medskip
Our results for $\kappa^{(2)}_{c,\,\rm dressed}$, as obtained from
Eq. (\ref{m2dressed}), are
listed in Table \ref{tab6}. 
 Comparing with the Monte Carlo estimates, dressed results show a
definite improvement over non-dressed values. It is interesting to
note that 1-loop dressed results already provide most of the
improvement, except at very large $\beta$-values. At the same time, a
sizeable discrepancy still remains, as was expected from start;
multiplicative renormalizations, calculated to the same order, are
expected to be much closer to their 
exact values. A first case study of this kind, regarding the
$\beta$-function with clover improvement, has now been completed and is
presented in Ref.~\cite{BP}.


\begin{table}[ht]
\begin{center}
\begin{minipage}{15cm}
\caption{Coefficients $c^{(1)}_{1,i}$, $c^{(1)}_{2,i}$,
$c^{(1)}_{3,i}$. $r=1$.
\label{tab1}}
\begin{tabular}{cr@{}lr@{}lr@{}l}
\multicolumn{1}{c}{$i$}&
\multicolumn{2}{c}{$c^{(1)}_{1,i}$} &
\multicolumn{2}{c}{$c^{(1)}_{2,i}$} &
\multicolumn{2}{c}{$c^{(1)}_{3,i}$} \\
\tableline \hline
4          & -0&.00697298969(1) &  0&.00711962270(1) &  0& \\
12+13      &  0&                &  0&                & -0&.00005540(1) \\
14+15+16+17+18&  0&.005587(1)   & -0&.0054356(3)     &  0& \\
19+20      &  0&                &  0&                & -0&.0004905(2)  \\
21+22+23   &  0&.0015499(6)     & -0&.0015499(6)     &  0& \\
24         & -0&.00022742(6)    &  0&                &  0& \\
25         &  0&.0014715(3)     & -0&.00002752(1)    &  0& \\
26         &  0&.0009439(2)     &  0&                &  0& \\
27         &  0&                & -0&.0007525(1)     &  0& \\ 
28         &  0&.00024887(3)    &  0&.000086138(5)   &  0& \\
\end{tabular}
\end{minipage}
\end{center}
\end{table}
\begin{table}[ht]
\begin{center}
\begin{minipage}{15cm}
\caption{Coefficients $c^{(2)}_{1,i}$, $c^{(2)}_{2,i}$,
$c^{(2)}_{3,i}$. $r=1$.
\label{tab2}}
\begin{tabular}{cr@{}lr@{}lr@{}l}
\multicolumn{1}{c}{$i$}&
\multicolumn{2}{c}{$c^{(2)}_{1,i}$} &
\multicolumn{2}{c}{$c^{(2)}_{2,i}$} &
\multicolumn{2}{c}{$c^{(2)}_{3,i}$} \\
\tableline \hline
4        & -0&.00486917062(1) &  0&.00452394220(1)     &  0&\\
6        &  0&.0017538(2)     &  0&                    &  0&\\
12+13    &  0&                &  0&                    &  0&.0008949(1) \\
14+15+16+17+18&  0&.0021977(2)& -0&.0022620(2)         &  0&\\
19+20    &  0&                &  0&                    &  0&.0004703(1) \\
21+22+23 & -0&.0001864(1)     &  0&.0001864(1)         &  0&\\
24       &  0&.00003257(1)    &  0&                    &  0&\\
25       & -0&.00022829(1)    & -0&.00005915(1)        &  0&\\
26       &  0&.00060875(1)    &  0&                    &  0&\\
27       &  0&                &  0&.00035168(2)        &  0&\\ 
28       &  0&.00053539(5)    & -0&.00011818(1)        &  0&\\
\end{tabular}
\end{minipage}
\end{center}
\end{table}
\begin{table}[ht]
\begin{center}
\begin{minipage}{15cm}
\caption{Coefficients $c^{(3)}_{1,i}$, $c^{(3)}_{2,i}$,
$c^{(3)}_{3,i}$. $r=1$.
\label{tab3}}
\begin{tabular}{cr@{}lr@{}lr@{}l}
\multicolumn{1}{c}{$i$}&
\multicolumn{2}{c}{$c^{(3)}_{1,i}$} &
\multicolumn{2}{c}{$c^{(3)}_{2,i}$} &
\multicolumn{2}{c}{$c^{(3)}_{3,i}$} \\
\tableline \hline
19+20   &  0&               &  0&               & -0&.00069225(3) \\
21+22+23& -0&.00017530(6)   &  0&.00017530(6)   &  0& \\
25      &  0&.000022090(4)  &  0&               &  0& \\
26      & -0&.000023954(3)  &  0&               &  0& \\
27      &  0&               & -0&.000017264(2)  &  0& \\ 
28      &  0&.000014014(3)  &  0&               &  0& \\
\end{tabular}
\end{minipage}
\end{center}
\end{table}
\begin{table}[ht]
\begin{center}
\begin{minipage}{15cm}
\caption{Coefficients $c^{(4)}_{1,i}$, $c^{(4)}_{2,i}$,
$c^{(4)}_{3,i}$. $r=1$.
\label{tab4}}
\begin{tabular}{cr@{}lr@{}lr@{}l}
\multicolumn{1}{c}{$i$}&
\multicolumn{2}{c}{$c^{(4)}_{1,i}$} &
\multicolumn{2}{c}{$c^{(4)}_{2,i}$} &
\multicolumn{2}{c}{$c^{(4)}_{3,i}$} \\
\tableline \hline
19+20    &  0&              &  0&               & -0&.00019810(1) \\
21+22+23 & -0&.000017219(2) &  0&.000017219(2)  &  0&             \\
27       &  0&              &  0&.000025610(2)  &  0&             \\ 
\end{tabular}
\end{minipage}
\end{center}
\end{table}
\begin{table}[ht]
\begin{center}
\begin{minipage}{15cm}
\caption{Coefficients $c^{(0)}_{1,i}$, $c^{(0)}_{2,i}$,
$c^{(0)}_{3,i}$. $r=1$.
\label{tab5}}
\begin{tabular}{cr@{}lr@{}lr@{}l}
\multicolumn{1}{c}{$i$}&
\multicolumn{2}{c}{$c^{(0)}_{1,i}$} &
\multicolumn{2}{c}{$c^{(0)}_{2,i}$} &
\multicolumn{2}{c}{$c^{(0)}_{3,i}$} \\
\tableline \hline
3    &  0&.002000362950707492 & -0&.0030005444260612375 &  0&\\
4    &  0&.00040921361(1)     & -0&.00061382041(2)      &  0&\\
6    & -0&.0000488891(8)      &  0&.000097778(2)        &  0&\\
7+8+9+10+11 & -0&.013927(3)   &  0&.014525(2)           &  0&\\
12+13 & 0&                    &  0&                     &  0&.00079263(8) \\
14+15+16+17+18 & -0&.005753(1)&  0&.0058323(7)          &  0&\\
19+20&  0 &                   &  0&                     &  0&.000393556(7) \\
21+22+23 & 0&.000096768(4)    & -0&.000096768(4)        &  0&\\
25   &  0&.00007762(1)        & -0&.00015524(3)         &  0&\\
26   & -0&.00040000(5)        &  0&                     &  0&\\
27   &  0&                    & -0&.000006522(1)        &  0&\\ 
28   &  0&.0000078482(5)      & -0&.000015696(1)        &  0&\\
\end{tabular}
\end{minipage}
\end{center}
\end{table}
\begin{table}[ht]
\begin{center}
\begin{minipage}{15cm}
\caption{One- and two-loop results ($\kappa_c^{(1)}$,
$\kappa_c^{(2)}$), along with their improved (dressed) counterparts,
and nonperturbative determinations. See references, shown in square
brackets, for details on the nonperturbative definition of $\kappa$
and on error estimates. \label{tab6}}
\medskip
\begin{tabular}{lllllllll}
\multicolumn{1}{c}{$N_f$}&
\multicolumn{1}{c}{$\beta$}&
\multicolumn{1}{c}{$c_{\rm SW}$}&
\multicolumn{1}{c}{$\kappa_c^{(1)}$}&
\multicolumn{1}{c}{$\kappa_c^{(2)}$}&
\multicolumn{1}{c}{$\kappa_{c,\,\rm dressed}^{(1)}$}&
\multicolumn{1}{c}{$\kappa_{c,\,\rm dressed}^{(2)}$}&
\multicolumn{2}{c}{Simulation}\\
\tableline \hline
0&5.70&1.568 &0.1296  &0.1332  &0.1366  &0.1366  &0.1432  &\cite{Bowler}\\
0&6.00&1.479 &0.1301  &0.1335  &0.1362  &0.1362  &0.1392  &\cite{Bowler}\\
0&6.00&1.769 &0.12749 &0.13061 &0.13372 &0.13319 &0.13525 &\cite{Bowler,LSSWW}\\
0&6.20&1.442 &0.1303  &0.1334  &0.1358  &0.1358  &0.1379  &\cite{Bowler}\\
0&6.20&1.614 &0.12878 &0.13182 &0.13439 &0.13414 &0.13582 &\cite{Bowler,LSSWW}\\
0&12.0&1.1637&0.128766&0.129622&0.129807&0.129845&0.129909&\cite{LSSWW}\\
0&24.0&1.0730&0.127019&0.127229&0.127243&0.127253&0.127258&\cite{LSSWW}\\
2&5.20&2.0171&0.12515 &0.12987 &0.13481 &0.13342 &0.13663 &\cite{UKQCD,JS}\\
2&2.26&1.9497&0.12589 &0.13043 &0.13517 &0.13392 &0.13709 &\cite{UKQCD,JS}\\
2&2.29&1.9192&0.12622 &0.13068 &0.13532 &0.13414 &0.13730 &\cite{UKQCD,JS}\\
\end{tabular}
\end{minipage}
\end{center}
\end{table}
\end{document}